\documentclass[prb,twocolumn,preprintnumbers,amsmath,amssymb,
footinbib,superscriptaddress]{revtex4-1}

\usepackage[titletoc,toc,title]{appendix}
\usepackage{braket}
\usepackage[final]{feynmp}
\usepackage{ifpdf}
\usepackage{comment}
\usepackage{amsmath}
\usepackage{mathrsfs}
\usepackage{color}
\usepackage{bm}

\usepackage[demo]{graphicx}
\usepackage{dcolumn}
\usepackage{bm}
\usepackage{hyperref}

\newcommand{\beq}{\begin{equation}}
\newcommand{\eeq}{\end{equation}}

\begin{document}

\title{Coulomb-induced instabilities of nodal surfaces}

\author{Pavel A. Volkov}
\affiliation{Theoretische Physik III, Ruhr-Universit\"{a}t Bochum, D-44780 Bochum, Germany}
\affiliation{Department of Physics and Astronomy, Center for Materials Theory, Rutgers University, Piscataway, NJ 08854 USA}
\author{Sergej Moroz}
\affiliation{Department of Physics, Technical University of Munich, 85748 Garching, Germany}

\begin{abstract}
We consider the stability of nodal surfaces in fermionic band systems with respect to the Coulomb repulsion. It is shown that nodal surfaces at the Fermi level are gapped out at low temperatures due to emergent particle-hole orders. Energy dispersion of the nodal surface suppresses the instability through an inhomogenous phase. We argue that around criticality the order parameter fluctuations can induce superconductivity. We show that by tuning doping and disorder one could access various phases, establishing fermionic nodal surface systems as a versatile platform to study emergent quantum orders.

\end{abstract}

\maketitle

%%%%%%%%%%%%%%%%%%%%%%%%%%%%%%%%%
%%%%%%%%%%%%%%%%%%%%%%%%%%%%%%%%%
{\it Introduction:}
The last decade saw a surge of research activity aimed at a better understanding of the physics of nodal objects in three-dimensional fermionic band structures \cite{Turner2013, Burkov2015, Armitage2018}. Most studied is the case of Weyl semimetals with nodal points, recently discovered experimentally\cite{Xu2015, Lv2015}. They generically appear in any band structure provided time-reversal or inversion symmetry is broken. Remarkably, nodal points do not exhaust all possible nodal objects in a three-dimensional band structure. In 2011 Weyl nodal loops, where two bands intersect each other on closed two-dimensional curves, were predicted \cite{Heikk2011, Heikkil2011a,  Burkov2011}. A generic nodal loop has to be protected by symmetries. It is distinguished by a $\pi$ Berry phase on a contour that links with it and exhibits nearly flat drumhead states in the boundary spectrum \cite{Burkov2011, Kim2015}.

More recently another type of nodal object was proposed - a nodal surface \cite{Liang2016, Zhong2016}. The nodal surface is a two-dimensional degeneracy of energy bands that forms a closed surface in the Brillouin zone. While less generic then the nodal points, they can nevertheless appear in systems that possess certain additional symmetries, such as sublattice or mirror symmetries. Nodal surfaces protected by different symmetries were theoretically predicted \cite{Agterberg2017, Timm2017, Bzdusek2017, Turker2018, Xiao2017, Wang2018} and their topological stability against small perturbations of the non-interacting Bloch Hamiltonian was investigated \cite{Bzdusek2017, Turker2018, Xiao2017}. Material realizations of nodal surfaces were also proposed \cite{Wu2018, Wang2018}.

However, to understand the robustness of nodal objects in real solids it is imperative to study their stability with respect to interactions. In Weyl point\cite{Zhong2013,Maciejko2014,Roy2017} and loop\cite{Shouvik2016} semimetals it has been shown that short-range interactions need to be stronger then a certain threshold to destroy the nodal structures. The long-range part of the Coulomb interaction is marginally irrelevant \cite{Hosur2012, Isobe2012} in the case of Weyl points.  In nodal line systems the  Coulomb interaction is partially screened, but  turns out to be irrelevant as well \cite{Huh2016}. On the other hand, introducing a finite density of states by breaking inversion symmetry \cite{Zyuzin2012} or doping \cite{Wang2016, Nandkishore2016} results in instabilities already at weak coupling.

In this paper we study the fate of nodal surfaces in the presence of the Coulomb interaction. First we show that the long-range part of the Coulomb interaction renders nodal surfaces at the Fermi level unstable to formation of an 'excitonic insulator', first proposed by Keldysh and Kopaev in the seminal paper \cite{Keldysh1965}. Subsequently,
we demonstrate that the energy dispersion of the nodal surface suppresses the instability. We specify our discussion to the case where the band structure has two-fold degeneracy in the Brillouin zone supplemented by spin rotational invariance, giving rise to the so-called Dirac nodal surface. We analyze the emerging orders in mean-field approximation for different combinations of short- and long-range repulsion and in the presence of sublattice symmetry. Additionally, we show that doping and disorder can serve as experimental 'knobs' that grant access to various phases including the particle-hole counterpart of the elusive FFLO state\cite{FF1964,LO1965} of superconductors. Finally, we argue that fluctuations of the particle-hole order parameter might lead to unconventional superconductivity in the vicinity of the doping-induced quantum critical point.

\begin{figure}[h!]
\centering
  \includegraphics[width=\linewidth]{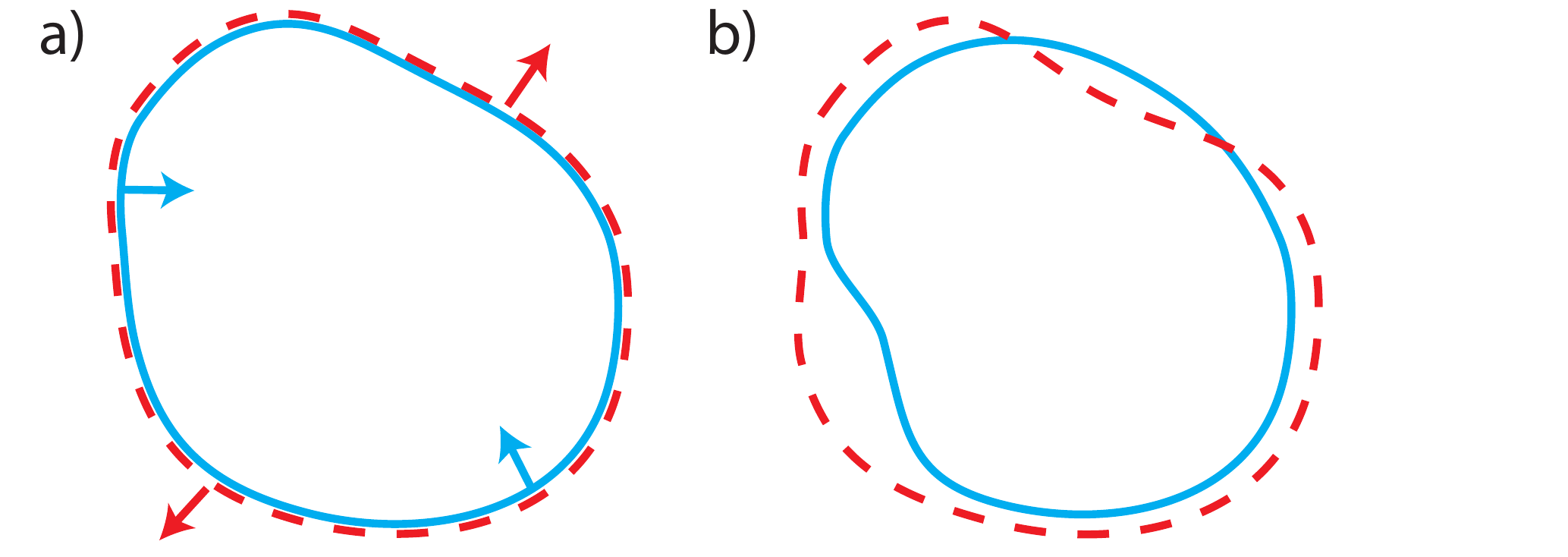}
  \caption{Two-dimensional projection of the electron (dashed red) and hole (solid blue) Fermi surfaces for a system with a nodal surface (\ref{eq:h0}) a) at zero energy; arrows show the direction of Fermi velocities b) with energy dispersion (see text); crossings of the two surfaces are projections of a single zero-energy nodal line. }
  \label{fig1}
\end{figure}

%%%%%%%%%%
{\it Particle-hole instability of the nodal surface:} We start by showing that long-range Coulomb repulsion can render nodal surfaces unstable. First we consider the case of a particle-hole symmetric Weyl nodal surface at the Fermi energy described by the two-band Hamiltonian
\begin{equation}
\hat{H}_0 =\sum_{\bf p}
\hat{c}_{\bf p}^\dagger
\varepsilon({\bf p})\sigma_z
\hat{c}_{\bf p},
\label{eq:h0}
\end{equation}
where the energy $\varepsilon({\bf p})$ is measured with respect to the Fermi level. In what follows, we will use the linearized form of the dispersion near the Fermi level $\varepsilon({\bf p})\approx {\bf v}_F(\theta, \varphi) \cdot ({\bf p}-{\bf p}_F(\theta, \varphi))$, where $\theta$ and $\varphi$ are the angles in spherical coordinates, assuming $|{\bf v}_F(\theta, \varphi)|\neq0$. Without interaction, the system described by (\ref{eq:h0}) contains an electron- and a hole-like Fermi surface that coincide in the momentum space (see Fig. \ref{fig1}a). For the following discussion we will consider the generalization of \eqref{eq:h0} to the case with $N_0$ identical nodal surfaces, which results in $N_0$ pairs of coinciding electron and a hole-like Fermi surfaces.

Importantly, the system has a finite density of itinerant charge carriers and the Coulomb potential is screened. In the random phase approximation (that can be justified in the regime $N_0\gg1$ \cite{Ye1999,Kharitonov2008}) one has
$4 \pi  e^2/ {\bf q}^2\to ({\bf q}^2 /4 \pi e^2 -\Pi(i \omega_n, {\bf q}))^{-1}$. To study the low-energy effects we approximate the interaction by its static ($\omega_n=0$) value at low momenta \footnote{For the analysis presented in this paper, relevant transfer momenta are of order of the Fermi momentum $p_F$, while the condition of small momenta reads $q\ll q_{TF}= \sqrt{4\pi e^2 \nu}$. As $\nu\sim N_0 p_F$, it is true either for a small $p_F$ or for $N_0\gg1$. Additionally, we assume weak coupling limit, such that in the logarithmic approximation it is sufficient to consider only static ($\omega_n=0$) screening \cite{Kharitonov2010}.}  such that $\Pi(i \omega_n, {\bf q})$ reduces to $-\nu$ ($\nu$ being the total density of states) resulting in
\begin{equation}
\hat{H}_{Coul} \approx \frac{U_0}{2\mathcal{V}} \sum_{{\bf p}, {\bf p}',{\bf q}}
\hat{c}_{{\bf p+q},\sigma}^\dagger  \hat{c}_{{\bf p'-q},\sigma'}^\dagger \hat{c}_{{\bf p'},\sigma'}   \hat{c}_{{\bf p},\sigma},
\label{eq:hint}
\end{equation}
where $\mathcal{V}$ is the volume of the system and $U_0=\nu^{-1}$. Note that $\nu=2 N_0 \nu_0$, where $\nu_0$ is the density of states of a single Fermi surface (electron- or hole-like).

Let us now consider the possible instabilities of (\ref{eq:h0}) with interaction (\ref{eq:hint}). Decoupling the repulsive interaction (\ref{eq:hint}) with the Hubbard-Stratonovich procedure one immediately finds that the particle-hole channels with order parameters $W_i({\bf Q}) = U_0\sum_{\bf p}\langle\hat{c}_{{\bf p}}^\dagger \sigma_i \hat{c}_{{\bf p}+{\bf Q}}\rangle$ are all attractive with the same coupling constant.
Moreover, the $W_x$ and $W_y$ orders for ${\bf Q} = {\bf 0}$ can be shown to develop a weak-coupling instability. The self-consistency equation for these orders has the form:
\begin{equation}
\frac{1}{U_0} =  \frac{\nu_0}{2}\int d\varepsilon
\frac{\tanh \frac{\sqrt{\varepsilon^2+W^2}}{2T}}{\sqrt{\varepsilon^2+W^2}},
\label{eq:wmf}
\end{equation}
where the density of states is approximated by a constant $\nu(\varepsilon)\approx\nu_0$ and $W=\sqrt{W_x^2+W_y^2}$. Due to the logarithmic divergence of the r.h.s. at low temperatures we see that a transition occurs for an arbitrary weak coupling with the critical temperature being
\begin{equation}
T_c = \frac{2 e^\gamma}{\pi} \Lambda e^{-2N_0},
\label{eq:tc}
\end{equation}
and the value of the order parameter at zero temperature $W_0 = 2 \Lambda e^{-2 N_0}$, where $\Lambda$ is the upper cutoff for the energy integral in (\ref{eq:wmf}). Physically, $\Lambda$ is determined by the structure of the dispersion in (\ref{eq:h0}) away from the Fermi surface. In particular, it is fixed by the bandwidth and the Fermi energy with respect to the band bottom. The eigenenergies in the presence of the order parameter are $E({\bf p})=\pm\sqrt{\varepsilon({\bf p})^2+W^2}$ clearly showing that a gap has opened and the nodal surface is destroyed.

For a single Weyl nodal surface ($N_0=1$), the transition corresponds to breaking of a global $U(1)_z$ symmetry generated by $\sum_{\bf p} \hat{c}_{\bf p}^\dagger \sigma_z \hat{c}_{\bf p}$. We note that the degeneracy between $W_x$ and $W_y$ orders is due to the highly symmetric form of the interaction (\ref{eq:hint}); additional interactions can break this degeneracy such that only a particular linear combination of them will develop a non-zero expectation value. However, if these additional interactions are much smaller then $U_0$ the equations for $T_c$ and $W_0$ will approximately hold. We discuss in detail the influence of additional interactions below for the case of a Dirac nodal surface.

{\it Nodal surfaces with energy dispersion:} While presence of nodal surfaces in the band structure can be guaranteed by symmetry and the ensuing topological invariants\cite{Bzdusek2017}, they do not have to be pinned to the Fermi energy. Indeed, a term of the form $f({\bf p})\sigma_0$ can be always added to the Hamiltonian (\ref{eq:h0}) without breaking any symmetries generated by $\sigma_i$. This term shifts the nodal surface $\varepsilon({\bf p})=0$ to an energy $f({\bf p})$ and breaks the particle-hole symmetry. As a result, a generic nodal surface coincides with Fermi surfaces on a set of loops in momentum space.

Physically, a constant $f({\bf p})=f_0$ is equivalent to a shift of the chemical potential. Consequently, a constant $f_0$ can be experimentally realized by doping electrons/holes into the system. Momentum-dependence of $f({\bf p})$, on the other hand, is determined by the specific material. For the particular case of a Dirac nodal surface protected by sublattice symmetry (see below), $f({\bf p})$ corresponds to intra-sublattice hopping.

We can now expand the function $f({\bf p})$ near the Fermi surface $f({\bf p}) \approx \gamma(\theta,\varphi)+\delta(\theta,\varphi) \varepsilon({\bf p}) +O(\varepsilon^2)$. The resulting eigenenergies of the free Hamiltonian are then $E_{\pm}=(\delta(\theta,\varphi)\pm 1) \varepsilon({\bf p}) + \gamma(\theta,\varphi)$. $\gamma(\theta,\varphi)$ leads to the electron and hole Fermi surfaces not being coincident anymore(see Fig. \ref{fig1}b), while $\delta(\theta, \varphi)$ gives rise to a difference between the Fermi velocities of the electron and hole bands. The latter can be shown \cite{supmat} not to destroy the weak-coupling instability.

Let us now analyze the physical consequences of a constant $\gamma(\theta,\varphi) = \gamma_0$. We assume $\gamma_0\ll E_F$ such that the density of states remains unchanged. In this case the mean-field Hamiltonian can be written as
\[H_{MF}=
\begin{bmatrix}
\hat{c}_{{\bf p},1}
\\
\hat{c}_{{\bf p},2}
\end{bmatrix}
^\dagger
\begin{bmatrix}
\varepsilon({\bf p})+\gamma_0 & W_x-i W_y\\
W_x+i W_y&-\varepsilon({\bf p})+\gamma_0
\end{bmatrix}
\begin{bmatrix}
\hat{c}_{{\bf p},1}
\\
\hat{c}_{{\bf p},2}
\end{bmatrix}
.
\]
One can note the similarity to the Hamiltonian of a superconductor in a Zeeman field. As is well known\cite{LO1965}, at low temperature the order parameter vanishes via a first-order transition at the Clogston-Chandrasekar limit $\gamma_0^{cr} = W_0/\sqrt{2}$, where $W_0$ is the order parameter in the absence of the Zeeman field. This implies that $\gamma_0$ stabilizes the nodal surface against weak enough interactions.

\begin{figure}[h!]
\centering
  \includegraphics[width=\linewidth]{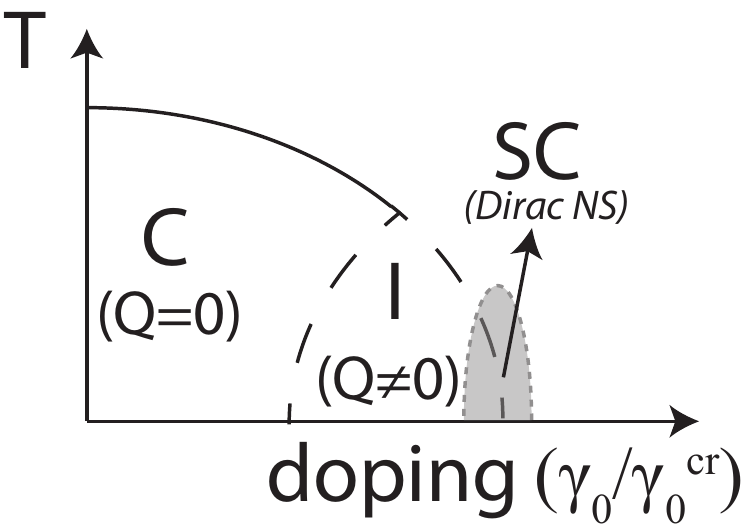}
  \caption{Tentative temperature-doping phase diagram for a nodal surface system with repulsive interaction. Doping scale is given by the value of $\gamma_0$ (see text) normalized  to $\gamma_0^{cr} = W_0/\sqrt{2}$. 'C' is the commensurate (homogenous) particle-hole order, while 'I' is the modulated phase. A superconducting ('SC') phase around the particle-hole QCP (shaded grey area) could emerge for the Dirac nodal surfaces.}
  \label{fig2}
\end{figure}

Importantly, the first order phase transition is preempted by an instability to the formation of a periodically modulated state \cite{LO1965,FF1964,Chevy2010}. In our case this corresponds to a particle-hole order parameter of the form $\hat{c}^\dagger_{\alpha,{\bf p+Q}}\hat{c}_{\alpha',{\bf p}}$, i.e. a charge-, spin-, or bond current density wave. A formal analogy extends also to itinerant antiferromagnets \cite{Rice1970,Cvetkovic2009,Vorontsov2009}. A qualitative phase diagram is presented in Fig. \ref{fig2}. As the details of the modulated (incommensurate) phase such as the direction and magnitude of the modulation wavevector are quite sensitive to the Fermi surface details, we do not consider them here. Generally, one would expect ${\bf Q}$ to connect the points with opposite Fermi velocities with the lowest curvature. Hence, low-dimensionality and presence of flat portions of the Fermi surface promote such phases\cite{Chevy2010}.

Let us move on to the effect of a non-constant $\gamma(\theta,\varphi)$. For the homogenous order, given the critical temperature $T_{c}$ for $\gamma=0$ \eqref{eq:tc}, one can extract the new critical temperature $T_c'$ from the equation \cite{supmat}
$
\log \frac{T_c'}{T_{c}} =
-\frac{1}{2}\langle
\psi\left(\frac{1}{2}+\frac{i|\gamma|}{2 \pi T_c'}\right)
+\psi\left(\frac{1}{2}-\frac{i|\gamma|}{2 \pi T_c'}\right)
-2\psi\left(\frac{1}{2}\right)
\rangle_{FS},
$
where $\psi(z)$ is the digamma function and $\langle...\rangle_{FS} =\int ... \frac{\nu_0(\varepsilon=0, \theta,\varphi)d\Omega}{4\pi\nu_0}$. We observe that the physical effect of $T_c$ suppression is present even if $\langle \gamma \rangle_{FS}=0$. At low  temperatures, unlike for $\gamma=$const, the order parameter may change its value before disappearing, since electron/hole pockets can start to appear in regions where $|\gamma(\theta,\varphi)|>W_0$. We note that for the incommensurate phases the results will depend on the particular realization of $\gamma({\bf p})$ and thus we shall not consider them here.

%%%%%

{\it Dirac nodal surface:}
Up to now, we have not discussed the physical meaning of the emergent orders $W_i$. To do so, we need to know how the fermionic operators $\hat{c}_{\bf p}$ change under symmetry operations, which, in general, depends on the system. Here we analyze the particular case of a single Dirac nodal surface ($N_0=2$) protected by inversion enriched time-reversal symmetry (TRS+I) and sublattice symmetry on a bipartite lattice (BDI class in \cite{Bzdusek2017}). In this case the system has full spin rotational invariance and the Hamiltonian retains the form (\ref{eq:h0}), being trivial (unity matrix) in spin space. On the other hand, in the sublattice basis the Hamiltonian has to be off-diagonal due to the sublattice symmetry. Moreover, the inversion-enriched TRS acts as complex conjugation (but does not flip the momentum) leading to $H({\bf p}) = H^*({\bf p})$. Consequently, the tight-binding Hamiltonian respecting the symmetry can contain only the real intersublattice hopping  $\sim \sigma_x$. Thus the unitary transformation $U=(\sigma_x+\sigma_z)/\sqrt{2}$, which acts as $U^{-1}\sigma_zU=\sigma_x$, transforms the operators from the 'band basis' of (\ref{eq:h0}) to the sublattice basis.

The instability analysis performed above can be directly applied to the Dirac nodal surface problem. However, apart from $W_x$ and $W_y$, spinful orders $W_{i,j}=U_0\sum_{\bf p}\langle\hat{c}_{{\bf p}}^\dagger \sigma_{i} s_j
\hat{c}_{{\bf p}}\rangle$, where $i=x,y$ and $s_j$ denotes the set of spin Pauli matrices, also develop weak coupling instabilities.

Now let us identify the physical meaning of the respective orders. Since $\sigma_x \overset{U}{\to}\sigma_z$, the $W_x$ order parameter has the meaning of an energy offset between the two sublattices. This leads in turn to a charge offset resulting in a charge density wave (CDW). Note that the CDW in this case breaks only the sublattice, but not translational symmetry. $\sigma_y\overset{U}{\to}-\sigma_y$ and thus in the sublattice basis the $W_y$ order introduces a nonzero average $i\langle \hat{c}_A^\dagger\hat{c}_B-\hat{c}_B^\dagger\hat{c}_A \rangle$ that implies a current flowing between different sublattice sites $A$ and $B$. Since in the ground state the total current between the sublattices should be zero, compensating inter-unit cell currents should be also present with their structure being determined by the form of $\varepsilon({\bf p})$. The CDW/bond current orders, discussed above, preserve/break TRS, respectively.
Turning now to spin orders, one observes that $W_{x,i}$ break both sublattice and time-reversal symmetry and essentially represent intra-unit cell antiferromagnetism (AFM). On the other hand, the orders $W_{y,i}$ do not break TRS and correspond to spin current order.

\begin{figure}[h!]
\centering
  \includegraphics[width=\linewidth]{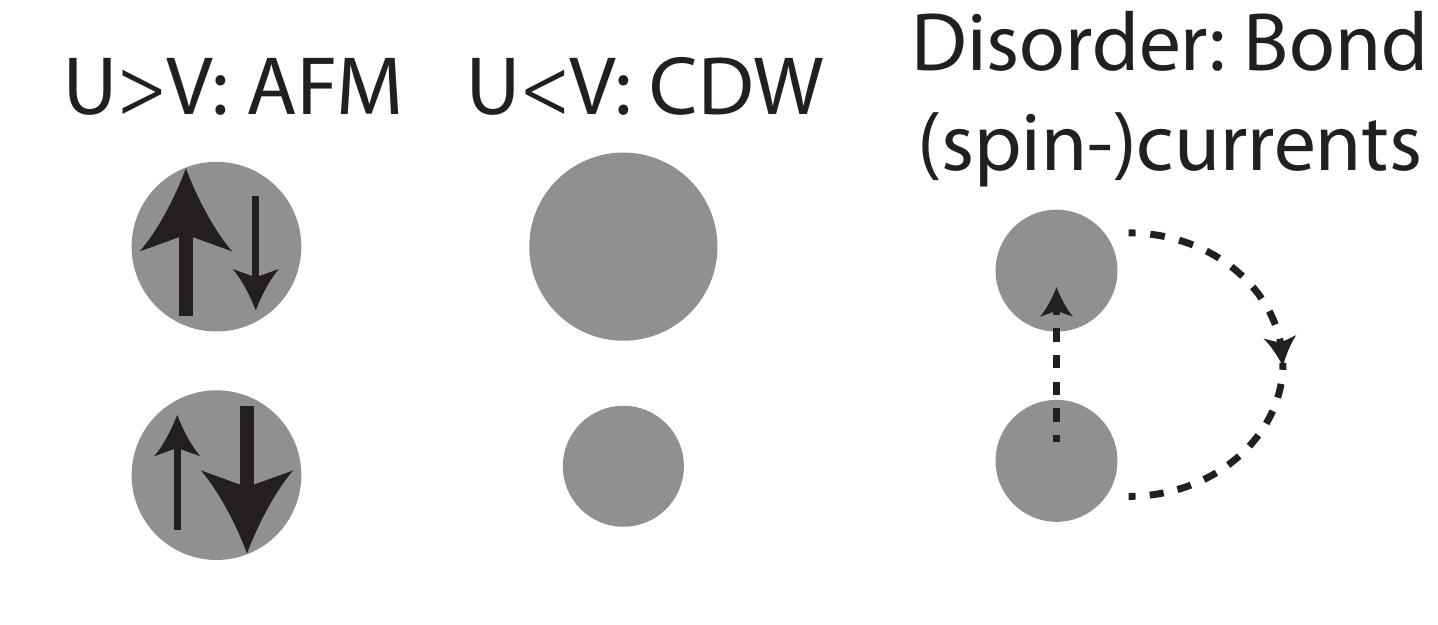}
  \caption{An illustration of the emergent states possible for a single Dirac nodal surface. Above the illustrations the necessary conditions for their realization are given.}
  \label{fig3}
\end{figure}

The interaction \eqref{eq:hint} results in the orders $W_{x,i}$ and $W_{y,i}$ forming a degenerate manifold. However, as the represented orders break different symmetries, one would expect this degeneracy to be accidental. Indeed, we show now that the inclusion of simplest on-site interactions allowed by symmetry lifts this degeneracy. Namely, let us add the on-site Hubbard repulsion $U [\hat n^A_{\uparrow} \hat n^A_{\downarrow}+\hat n^B_{\uparrow} \hat n^B_{\downarrow}]$ and the repulsion between the two sublattices $V \hat n^A \hat n^B$, where we introduced the densities $\hat n^{A(B)} = \hat c^\dagger(1\pm\sigma_x) \hat c/2$. Decoupling these interactions\cite{supmat} with respect to the same channels as before we get corrections to the transition temperatures \eqref{eq:tc} resulting from $2 N_0\to(1/(2 N_0)-\mathcal{V}_0\nu_0\lambda_{i,j})^{-1}$, where $\mathcal{V}_0$ is the unit cell volume. The coupling constants for the orders $W_{i,j}$ are
$
\lambda_{x,0} = U-2V$,
$\lambda_{x,\{x,y,z\}} = -U$  and $ \lambda_{y,\{0,x,y,z\}} = -V$.
Consequently, for $U>V$ the intra-unit cell antiferromagnetic order is the leading one, while for $U<V$ the CDW order wins (see Fig. \ref{fig3}). Note that the influence of the short-range interactions can be controlled by the density of states $\nu_0$, i.e. for dilute systems the screened Coulomb interaction \eqref{eq:hint} is still the dominant one. On the other hand, we consider $\lambda_{i,j}$ to be not too small, such that the fluctuations of competing orders could be neglected.

%%%%%%%%%%%%
{\it Effects of disorder:} As nodal surface systems are 3D metals with finite density of states, weak disorder that preserves all symmetries is not expected to disrupt nodal surfaces\cite{Lee1985,Syzranov2018}. On the other hand, interaction-induced particle-hole orders can be suppressed even for low impurity concentration. We show below that this effect allows one to promote the (otherwise subleading) staggered current phase discussed above (see Fig. \ref{fig3}).

We perform the calculation in the framework of Abrikosov-Gor'kov theory\cite{agdbook}. Namely, let us consider impurity potential affecting atoms of one of the sublattices only, corresponding to a single-site substitution or vacancy, with the Hamiltonian $\hat H_{imp}= u/\sqrt{2} \hat c^\dagger({\bf r}_0)(1\pm\sigma_x)\otimes s_0\hat c({\bf r}_0)$ (for other types of impurities see \cite{supmat}). Assuming randomly distributed impurity positions one obtains the equation for the critical temperature $T_c^d$ in the presence of disorder for the case of weak dilute impurities ($u p_F^3\ll E_F$, $p_F l\gg1$, where $l$ is the mean free path) for the orders $W_{x,i}$ and $W_{y,i}$
\begin{equation}
\begin{gathered}
\log \frac{T_c}{T_c^d} = \psi\left(\frac{1}{2}+\frac{\Gamma_{x,y}}{2 \pi T_c^d}\right)
-\psi\left(\frac{1}{2}\right)
,
\end{gathered}
\label{eq:dis:sigma_full}
\end{equation}
where $\Gamma_{x} = 4\Gamma,\; \Gamma_{y} = 2\Gamma$ with $\Gamma = \pi u^2 n \nu_0$, where $n$ is the impurity concentration. One can see the suppression rates are smaller for the $W_{y,i}$ orders, that have been identified above as subleading ($T_c^y<T_c^x$). The transitions will be completely suppressed for $\Gamma_{cr}^x = \frac{\pi}{8 e^\gamma} T_c^x$ and $\Gamma_{cr}^y = \frac{\pi}{4 e^\gamma} T_c^y$, for the $W_{x,i}$ and $W_{y,i}$ orders, respectively. Thus, if $T_c^y/T_c^x>1/2$, there exists a range of impurity concentrations such that the (spin-) current order can overcome the competing CDW or AFM order.

{\it Quantum critical point and unconventional superconductivity}: Above we pointed out the possibility of suppressing the particle-hole orders by doping (see also Fig.\ref{fig1}). With the situation being similar to the high-T$_c$ superconductors\cite{Hirschfeld2011,Scalapino2012}, it is natural to ask whether the fluctuations of the suppressed order around the critical doping can induce superconductivity. The interfermion interaction due to the critical fluctuations of a particle-hole order can be described by an effective action
\begin{equation}
S_{QCP} = -\frac{g T}{\mathcal{V}} \sum_{p, p', q}\chi(q) c^\dagger_{p+q} \hat{W} c_{p}
c^\dagger_{p'-q}\hat{W}c_{p'},
\label{eq:lqcp}
\end{equation}
where $\hat{W}$ is a matrix in the band and spin space corresponding to the particle-hole order parameter and $q\equiv({\bf q},\omega_n)$. $\chi(q)$ has the Ornstein-Zernike form $(\omega_n^2/c^2+({\bf q} - {\bf Q})^2+\xi^{-2})^{-1}$, $\xi$ being the correlation length of the fluctuations. Assuming $Q\ll p_F$ and not too small $\xi$ we can restrict our consideration to momentum-independent order parameters $\Delta_{\alpha,\alpha'}=\sum_{\bf p}\hat{c}_{p,\alpha}\hat{c}_{-p,\alpha'}$. The condition for the interaction to be attractive is
$
{\rm Tr} [\Delta^\dagger \hat{W} \Delta \hat{W}^T]>0.
$
Moreover, only intraband $\sim\sigma_{0,z}$ pairing results in logarithmic enhancement at low temperatures. For CDW QCP $\hat W\sim\sigma_x$ and conventional singlet superconductivity $\Delta \sim\sigma_0\otimes is_y$ is promoted. For AF QCP, on the other hand, $\hat W\sim\sigma_x\otimes \vec{s}$ and the singlet channel $\Delta \sim\sigma_z\otimes is_y$ is attractive. This type of pairing corresponds to a full-gap superconductivity with a sign change between the electron and hole Fermi surfaces similar to the $s_{\pm}$ state in the iron-based superconductors\cite{Hirschfeld2011}.

{\it Outlook and conclusion:} Our results can be readily applied to a number of proposed physical realizations of nodal surfaces \cite{Zhong2016,Wu2018, Wang2018}. For systems with few small nodal surfaces, such as $YH_3$ proposed in \cite{Wang2018}, one can expect that the long-range part of the Coulomb repulsion is likely to be the dominant interaction. On the other hand, in the case of graphene networks\cite{Zhong2016} one has two Dirac nodal surfaces ($N_0=4$) and thus the instability is likely to be driven by short-range repulsion or electron-phonon interactions not considered here. Finally, we note that the layered material ZrSiS is a promising candidate for hosting weakly dispersive nodal surfaces, with dispersion provided by the interlayer hopping. Recently it has been studied in \cite{Rudenko2018} using  a 2D square lattice model with a nodal line.

Overall, our results show that systems with nodal surfaces could serve as a potential platform for realizing a multitude of quantum orders (Fig.\ref{fig2}, Fig.\ref{fig3}) with many experimentally accessible 'knobs', such as doping or disorder, to probe the phase diagram.

\begin{acknowledgments}
We acknowledge useful discussions with Tom\'a\v s Bzdu\v sek. The work of S.M. is supported by the Emmy Noether Programme of German Research Foundation (DFG) under grant No. MO 3013/1-1. P.A.V. acknowledges the support by the Rutgers University Center for Materials Theory Postdoctoral fellowhsip.
\end{acknowledgments}

\bibliography{library}

\clearpage
\begin{widetext}
\section{Supplemental material: Coulomb-induced instabilities of nodal surfaces}
\setcounter{page}{1}
\renewcommand{\theequation}{S\arabic{equation}}
\setcounter{equation}{0}
\renewcommand{\thefigure}{S\arabic{figure}}
\setcounter{figure}{0}

\section{Critical temperature in presence of non-constant $\gamma(\theta,\varphi)$}
In the presence of $\gamma(\theta,\varphi)$ the self-consistent gap equation is
\beq
\frac{1}{U_0}=T\sum_{\varepsilon_n}\int d\varepsilon \frac{1}{(\varepsilon_n+i\gamma (\theta,\varphi))^2+W^2+\varepsilon^2}.
\eeq
The new critical temperature $T_c'$ can be computed as follows

\begin{gather*}
\log \frac{T_c'}{T_{c}} =
\int \frac{d\Omega}{4\pi}  T_c'\sum_{\varepsilon_n}\int d\varepsilon
\frac{1}{(\varepsilon_n+i\gamma (\theta,\varphi))^2+\varepsilon^2}-
\sum_{n=-\Lambda/(2\pi T_{c}')}^{\Lambda/(2\pi  T_{c}')} \frac{1}{|2n+1|}
\\
\approx
\int \frac{d\Omega}{4\pi} \pi  T_c'\sum_{\varepsilon_n}
\frac{{\rm sgn}\varepsilon_n}{\varepsilon_n+i\gamma(\theta,\varphi)}
-
\frac{1}{|\varepsilon_n|}
=
\int \frac{d\Omega}{4\pi}
\pi  T_c'\sum_{\varepsilon_n>0} \frac{-2\gamma^2(\theta,\varphi)}{|\varepsilon_n|(\varepsilon_n^2+\gamma^2(\theta,\varphi))}
=
\\
-\frac{1}{2}\langle
\psi\left(\frac{1}{2}+\frac{i|\gamma|}{2 \pi T_c'}\right)
+\psi\left(\frac{1}{2}-\frac{i|\gamma|}{2 \pi  T_c'}\right)
-2\psi\left(\frac{1}{2}\right)
\rangle_{FS},
\end{gather*}

where we assumed the cutoff energy to be much larger then $ T_c'$. Assuming $\gamma \ll  T_c'$ we get
\[
T_c'=T_{c}\left(1+\frac{\psi''(1/2)}{8\pi^2}\frac{\langle \gamma^2\rangle_{FS}}{T_{c}^2}\right).
\]

\section{Gap equation modification due to $\delta(\theta, \varphi)$}
Near the critical temperature the right-hand-side of the gap equation (3) %\eqref{eq:wmf}
should be replaced with
\[
\nu_0 \int d\varepsilon \frac{\tanh \frac{\varepsilon}{2T_c}}{ 2\varepsilon} \to \nu_0
\int d\varepsilon
\frac{\left\langle\tanh \frac{(1+\delta)\varepsilon}{2T_c}+\tanh \frac{(1-\delta)\varepsilon}{2T_c}\right\rangle_{FS}}{4\varepsilon}
.
\]
Splitting the integral into two and introducing the variables $\frac{(1\pm\delta)\varepsilon}{2T_c}$ in the respective integrals, we find that the only effect of $\delta$ is to renormalize the cutoff. For small $\delta$ one gets, e.g., $\Lambda\to\Lambda(1-\langle \delta^2\rangle_{FS})$. Thus one can see that $\delta$ does not destroy the weak-coupling instability.

%%%%%%%%%%%%%%%%%%%%%%%%%%%%%%%%%%%%
\section{Hubbard-Stratonovich procedure for generic interactions}
Let us consider a general short-range interaction term with a certain structure in the space of internal degrees of freedom
\[
g_{ij}\sum_{{\bf p}, {\bf p}',{\bf q}}
\hat{c}_{{\bf p+q}}^\dagger \Gamma_i\hat{c}_{{\bf p}}
\hat{c}_{{\bf p'-q}}^\dagger \Gamma_j  \hat{c}_{{\bf p'}}.
\]
For the mean-field analysis we need to perform the Hubbard-Stratonovich decoupling with respect to particular combinations of the fields $\hat{c}_{{\bf p},\alpha}^\dagger \hat{c}_{{\bf p},\beta}$. Let us denote $\sum_{\bf p}\hat{c}_{{\bf p},\alpha}^\dagger \hat{c}_{{\bf p},\beta}\equiv b_{\alpha\beta}$. Out of the total interaction we can extract two terms (by fixing one of the momenta) that can be decoupled
\[
H_{H}=\sum_{{\bf p}, {\bf p}'}
\hat{c}_{{\bf p}}^\dagger \Gamma_i\hat{c}_{{\bf p}}
\hat{c}_{{\bf p'}}^\dagger \Gamma_j  \hat{c}_{{\bf p'}},
\;
\text{and}
\;
H_{F}=
\sum_{{\bf p}, {\bf p}'}
\hat{c}_{{\bf p}'}^\dagger \Gamma_i\hat{c}_{{\bf p}}
\hat{c}_{{\bf p}}^\dagger \Gamma_j  \hat{c}_{{\bf p'}},
\]
where the first term corresponds to the Hartree-like energy and the second one to the Fock-like exchange. Now let us rewrite these expressions using $ b_{\alpha\beta}$
\[
H_{H}= g_{ij} {\rm Tr}\{ b \Gamma_i^t\} {\rm Tr} \{b \Gamma_j^t\},\quad
H_{F}= - g_{ij} {\rm Tr}\{\Gamma_j^t b \Gamma_i^t b\},
\]
where the superscript $t$ denotes transposition.
Decomposing $b$ into different channels $b=\frac{1}{2N_0}\sum_{l} b_l \Gamma_l$, where $b_l =  {\rm Tr} \{b \Gamma_l\}$ we can rewrite the total interaction term as
\begin{equation}
\begin{gathered}
H_{int} = \frac{1}{4 N_0}\sum_{lm}b_l b_m \lambda_{lm},
\\
\lambda_{lm} = \frac{1}{N_0} \sum_{ij} g_{ij} [ {\rm Tr}\{ \Gamma_l \Gamma_i^t\} {\rm Tr} \{\Gamma_m \Gamma_j^t\}\;
- {\rm Tr}\{\Gamma_j^t \Gamma_l \Gamma_i^t \Gamma_m\}].
\end{gathered}
\label{eq:hs_gen}
\end{equation}

The eigenvalues of the matrix $\lambda_{lm}$ give the interaction constants for the channels represented by the eigenvectors. The Hubbard-Stratonovich decoupling  is then performed for each channel separately.
In the case of the Dirac nodal surface, it is convenient to use the matrices $\sigma_a\otimes s_b$ as the basis $\Gamma$. In this representation the matrix $\lambda$ in (\ref{eq:hs_gen}) has four indices, i.e.,  $\lambda_{ab,cd}$. In the main text we discuss particular examples. The on-site Hubbard repulsion has the form
\begin{gather*}
U [n^A_{\uparrow}n^A_{\downarrow}+n^B_{\uparrow}n^B_{\downarrow}] =
\frac{U}{4} [c_{\uparrow}^\dagger(1+\sigma_x)c_{\uparrow}c_{\downarrow}^\dagger(1+\sigma_x)c_{\downarrow}
+c_{\uparrow}^\dagger(1-\sigma_x)c_{\uparrow}c_{\downarrow}^\dagger(1-\sigma_x)c_{\downarrow}]=
\\
\frac{U}{2} [c_{\uparrow}^\dagger c_{\uparrow} c_{\downarrow}^\dagger c_{\downarrow}
+ c_{\uparrow}^\dagger\sigma_x c_{\uparrow} c_{\downarrow}^\dagger\sigma_x c_{\downarrow}]=
\frac{U}{8} [ c^\dagger(1+s_z) c c^\dagger(1-s_z) c
+ c^\dagger\sigma_x(1+s_z) c_{\uparrow} c^\dagger\sigma_x(1-s_z) c]=
\\
\frac{U}{8} [( c^\dagger c)^2-( c^\dagger s_z c)^2
+( c^\dagger\sigma_x c)^2-( c^\dagger \sigma_x s_z c)^2].
\end{gather*}
The inter-sublattice repulsion is
\begin{gather*}
V n^An^B =\frac{V}{4} [ c^\dagger(1+\sigma_x) c c^\dagger(1-\sigma_x) c]=
\frac{V}{4} [( c^\dagger c)^2 - ( c^\dagger\sigma_x c)^2].
\end{gather*}
In the momentum space $U,V\to (U,V)/N$, where $N$ is the number of unit cells that can be expressed as $\mathcal{V}/\mathcal{V}_0$. A direct calculation using \eqref{eq:hs_gen} leads then to the results in the main text. The screened long-range part of the Coulomb repulsion yield $\lambda_{ll} = -U_0$.  As $\lambda_{l\neq m} = 0$ the channels do not mix and considering one particular channel one gets the following effective Lagrangian:
\[
L_{eff} =
\frac{N_0 \mathcal{V} W_l^2}{U_0-\lambda_l \mathcal{V}_0}+
\sum_{\bf p}
c^\dagger({\bf p},\tau) (\partial_\tau+\varepsilon({\bf p})\sigma_z+ W_l\Gamma_l) c({\bf p},\tau).
\]
Integrating the fermions out one can derive a self-consistency equation for $W_l$:
\[
\frac{2 W_l}{U_0-\lambda_l \mathcal{V}_0} = \frac{1}{N_0\mathcal{V}} T \sum_{\varepsilon_n, {\bf k}} \frac{\partial}{\partial W_l}
{\rm Tr}\log[i\varepsilon_n - \varepsilon({\bf p})\sigma_z- W_l\Gamma_l].
\]
For the case of $N_0$ Weyl nodal surfaces with intrasurface $W_{x/y}$ order this results in $\frac{1}{U_0-\lambda_l \mathcal{V}_0} = \nu_0\int d\varepsilon T \sum_{\varepsilon_n} \frac{1}{\varepsilon_n^2+\varepsilon^2+W^2}$.

\section{Effect of weak disorder on particle-hole instabilities}
We introduce a matrix Green's functions $-\langle T_\tau \psi({\bf p},\tau)_{\alpha} \overline{\psi} ({\bf p},0)_{\beta}\rangle$ with the anomalous part due to a particle-hole order parameter. As the order parameters considered have the off-diagonal structure with $\sigma_x$ or $\sigma_y$ we get

\[
G_0(i\varepsilon_n,{\bf p}) = (i\varepsilon_n -  \varepsilon({\bf p})\sigma_3 - \hat{W})=- \frac{i \varepsilon_n + \varepsilon({\bf p})\sigma_3 + \hat{W}}{\varepsilon_n^2+\varepsilon({\bf p})^2+W^2},
\]
where $\hat W$ corresponds to a particular order: $\hat W=W\sigma_{x,y}\otimes s_{0,x,y,z}$. The effect of impurities of a certain type at position are described by the potential
\[
\int d {\bf r} V_{imp}({\bf r})  c^\dagger({\bf r})\hat{T}c({\bf r}),
\quad
V_{imp}({\bf r}) =  u_T\sum_i \delta({\bf r}-{\bf r}_i)
\]
where the hermitian matrix $\hat{T}$ describes the intra-unit cell structure of the impurity. Subsequently, an averaging over the impurities positions is carried out such that
\[ \langle V_{imp}({\bf r})\rangle_{r_i} = n_T u_T,\quad \langle V_{imp}({\bf r})V_{imp}({\bf r}')\rangle_{r_i} = n_T u_T^2 \delta ({\bf r}-{\bf r}'), \]
where $n_T$ is the impurity concentration.
We assume that the impurities do not break the symmetries of the system on the average. First, we consider spinless impurities with $\hat{T}_0 = \sigma_0\otimes s_0$ and $\hat{T}_{\pm} = \frac{1}{\sqrt{2}}(1\pm\sigma_x)\otimes s_0$, the latter corresponding to a single-site substitution or vacancy (note that $u_+=u_-$ and the concentrations of $'+'$ and $'-'$ impurities are equal to preserve the sublattice symmetry). For paramagnetic impurities we write the impurity matrix as $\vec{S}\hat{T}\otimes\vec{s}$ (where $\hat T$ stands for $\hat T_0$ or $\hat T_{\pm}$ defined above) and subsequently average over the direction of $\vec{S}$: $\langle\vec{S}\rangle=0,\;\langle S_iS_j\rangle =\frac{S(S+1)}{3} \delta_{ij}$. The impurities introduced above act on a given lattice site. On the other hand, impurities that contain $\sigma_z$ or $\sigma_y$ correspond physically to a change of the hopping integrals. This effect should be considerably weaker than the potential directly induced by the charge or spin of an impurity at lattice sites and thus we do not consider them in detail.

Self-energy in the lowest order is simply equal to $u_0n_0+\sqrt{2}u_\pm n_\pm$ that should be compensated by an appropriate change of the chemical potential as we assume that impurities do not change the doping. In the second-order perturbation theory the effect of a particular type of impurity leads to the self-energy
\begin{equation}
\begin{gathered}
\Sigma(i \varepsilon_n) =
-\frac{\Gamma_i}{\pi} \int d \varepsilon \hat{T} \frac{i \varepsilon_n + \varepsilon\sigma_3 + \hat{W}}{\varepsilon_n^2+\varepsilon^2+W^2} \hat{T},
\end{gathered}
\label{eq:dis:sigma_1}
\end{equation}
where $\Gamma_i =\pi u_i^2 n_i \nu_0$ for nonmagnetic and $\Gamma_i^M = \pi(u_i^M)^2 n \nu_0 S(S+1)$ for magnetic impurities, where $i=0,+,-$.
The effect of this self-energy is to renormalize the $\varepsilon_n$ and $\hat{W}$ terms in the bare propagator: $\varepsilon_n\to \widetilde{\varepsilon_n}$ and $\hat{W} \to \widetilde{W} (\varepsilon_n)$. Consequently one can actually evaluate the sum of all non-crossing diagrams by evaluating (\ref{eq:dis:sigma_1}) with the renormalized $\varepsilon_n$ and $\hat{W}$. The resulting self-consistency equation (Dyson equation) is:
\begin{equation}
\begin{gathered}
G^{-1} + \Sigma = G_0^{-1},
\\
\widetilde{\varepsilon_n}\left(1-\Gamma_{\Sigma} \frac{1}{\sqrt{\widetilde{\varepsilon}_n^2+\widetilde{W}^2}}\right) = \varepsilon_n ,
\\
\widetilde{W}+\sum_i \Gamma_i
\frac{\hat{T}_i\widetilde{W}\hat{T}_i}
{
\sqrt{
\widetilde{\varepsilon}_n^2+\widetilde{W}^2
}
}
= \hat{W},
\end{gathered}
\label{eq:dis:sigma_full}
\end{equation}
where $\Gamma_{\Sigma} = \sum_i \Gamma_i= \Gamma_0+2\Gamma_\pm +\Gamma_0^M+2\Gamma_\pm^M$, the sum being over the considered types of impurities. Assuming $\widetilde{W}\sim\hat{W}\sim \sigma_{x,y}\otimes s_i$, one can rewrite the second one as
\begin{gather*}
\widetilde{W} = \hat{W} \left(1+\frac{\Gamma_W'}{\sqrt{\widetilde{\varepsilon}_n^2+\widetilde{W}^2}}\right)^{-1},
\end{gather*}
where $\Gamma'_{W}$ stands for
\begin{gather*}
\Gamma_{x,0}' = \Gamma_0+2\Gamma_\pm +\Gamma_0^M+2\Gamma_\pm^M,
\quad
\Gamma_{y,0}' = \Gamma_0+\Gamma_0^M,
\\
\Gamma_{x,\{x,y,z\}}' = \Gamma_0+2\Gamma_\pm -\frac{\Gamma_0^M+2\Gamma_\pm^M}{3},
\quad
\Gamma_{y,\{x,y,z\}}' = \Gamma_0-\frac{\Gamma_0^M}{3}.
\end{gather*}
We now introduce the renormalized Green's function in the self-consistency equation for the order parameter
\[
\frac{\hat{W}}{\nu_0\lambda} = T\sum_{\varepsilon_n} \frac{\pi\widetilde{W}}{\sqrt{\widetilde{\varepsilon}_n^2+\widetilde{W}^2}}.
\]
Near $T_c$ we can expand this equation along with ($\ref{eq:dis:sigma_full}$) to obtain
\[
1 = 2\pi\nu_0\lambda T \sum_{\varepsilon_n>0} \frac{1}{\varepsilon_n+(\Gamma_{\Sigma}+\Gamma_W')} \approx
\nu_0\lambda \left(\log\frac{\Lambda}{2\pi T_c^d}-\psi\left(\frac{1}{2} + \frac{\Gamma_{\Sigma}+\Gamma_W'}{2\pi T}\right)\right).
\]
It follows then that
\[
\log \frac{T_c}{T_c^d} = \psi\left(\frac{1}{2}+\frac{\Gamma_W}{2 \pi T_c^d}\right)
-\psi\left(\frac{1}{2}\right),
\]
where $\Gamma_W=\Gamma_{\Sigma}+\Gamma_W'$.
\end{widetext}

\end{document}